\documentclass{aa}  
\usepackage{graphicx,aalongtable}
\usepackage{txfonts}
\usepackage{natbib}
\bibpunct{(}{)}{;}{a}{}{,}
\begin{document}
   \title{Chemical abundances for the transiting planet host stars OGLE-TR-10, 
          56, 111, 113, 132 and TrES-1\thanks{Based on observations collected 
	  at the ESO 8.2-m VLT-UT2 Kueyen telescope (programs 075.C-0185 
	  and 076.C-0131).}}

   \subtitle{Abundances in different galactic populations}


   \author{N.C.~Santos\inst{1,2,3} \and
   	   A.~Ecuvillon\inst{4} \and
	   G. Israelian\inst{4} \and
	   M.~Mayor\inst{2} \and
           C.~Melo\inst{5} \and
	   D.~Queloz\inst{2} \and 
	   S.~Udry\inst{2} \and
	   J.P.~Ribeiro\inst{1} \and
	   S.~Jorge\inst{1}
           }

   \offprints{N.C. Santos, \email{nuno@oal.ul.pt}}

   \institute{
             Centro de Astronomia e Astrof{\'\i}sica da Universidade de Lisboa,
             Observat\'orio Astron\'omico de Lisboa, Tapada da Ajuda, 1349-018
             Lisboa, Portugal
	 \and    
             Observatoire de Gen\`eve, 
	     51 ch.  des Maillettes, CH--1290 Sauverny, Switzerland
	 \and 
	     Centro de Geofisica de \'Evora, Rua Rom\~ao Ramalho 59, 
	     7002-554 \'Evora, Portugal     
	 \and
             Instituto de Astrof{\'\i}sica de Canarias, 
	     E-38200 La Laguna, Tenerife, Spain
         \and
	     European Southern Observatory, 
	     Casilla 19001, Santiago 19, Chile
            }


   \date{Accepted for publication in Astronomy \& Astrophysics (June 2006)}

 
  \abstract
   {}
   {We used the UVES spectrograph (VLT-UT2 telescope) to obtain high-resolution 
    spectra of 6 stars hosting transiting planets, namely for OGLE-TR-10, 
    56, 111, 113, 132 and TrES-1. The spectra
    are now used to derive and discuss the chemical abundances 
    for C, O, Na, Mg, Al, Si, S, Ca, Sc, Ti, V, Cr, Mn, Co, Ni, Cu and Zn.}
   {Abundances were derived in LTE, using 1-D plane-parallel 
   Kurucz model atmospheres. For S, Zn and Cu we used a spectral synthesis 
   procedure, while for the remaining cases the abundances were derived 
   from measurements of line-equivalent widths. }
   {The resulting abundances are compared with those found for stars
   in the solar neighborhood. Distances and galactic coordinates are estimated
   for the stars. We conclude that besides being particularly metal-rich, 
    with small possible exceptions OGLE-TR-10, 
    56, 111, 113, 132 and TrES-1 are chemically undistinguishable from
    the field (thin disk) stars regarding their [X/Fe] abundances. This is particularly 
    relevant for the most distant of the targets, located at up to $\sim$2\,Kpc from the Sun. We also 
    did not find any correlation between the abundances and the condensation temperature of the
    elements, an evidence that strong accretion of planetary-like material,
    tentatively connected to planetary migration, did not occur. }
   {}


\maketitle

\section{Introduction}

The discovery of several short period transiting giant extra-solar
planets is giving astronomers the possibility to measure physical
variables like the planetary radius, mass and mean density.
In all, 10 transiting cases have been discovered. While some of
them are the outcome of the radial-velocity surveys \citep[][]{Charbonneau-2000,Henry-2000,Sato-2005,Bouchy-2005b},
a few others were discovered in the context of photometric transit searches
\citep[][]{Konacki-2003,Bouchy-2004,Pont-2004,Alonso-2004,Bouchy-2005a,Konacki-2005,McCullough-2006}.

The most prolific search for planetary transits has been
carried out by the OGLE team \citep[e.g.][]{Udalski-2002}. After follow-up
radial-velocity obervations, 5 of the more than 170 initial candidates
were confirmed as real planetary transits. Unfortunately, all the
OGLE transiting planets orbit faint (V$\sim$16) distant stars ($>$500\,pc), 
making it difficult the task of deriving accurate stellar parameters and chemical 
abundances.

Recently we have obtained high-resolution and high signal-to-noise
spectra of the 5 confirmed OGLE planet host stars using the UVES
spectrograph at the 8.2-m VLT/Kueyen telescope (ESO, Chile). 
These spectra were used to derive accurate stellar parameters for
the stars, which were then used to improve the estimates of the
planetary radii and mean-densities. The results for OGLE-TR-10, 56, 111 and 
113, as well as for the brighter TrES-1, were published in \citet[][]{Santos-2006a}. 
For OGLE-TR-132 the analysis was presented in Pont et al. (2006, in prep.).
In a separate paper \citep[][]{Melo-2006} we have further discussed the 
ages of these stars, showing that they are older than $\sim$0.5-2\,Gyr.
This conclusion has a strong impact on the evaporation rates of hot
and very-hot-jupiters. Finally, in \citet[][]{Guillot-2006} we have shown that
a correlation between the stellar metallicity and the planetary internal 
structure seems to exist. If real, this result will have a strong impact
on the models of planet formation.

In the current paper we present the abundances for several chemical elements
in the 6 stars mentioned above (OGLE-TR-10, 56, 111, 
113, 132 and TrES-1). Using the derived stellar parameters and observed
magnitudes, together with an estimate for the magnitude extinction, 
we then derive the distances to the targets and their galactic coordinates. 
The results are then discussed and the abundances
compared with those found for solar-neighborhood disk dwarfs.
Finally, we compare the abundances of volatile and refractory elements in
the 6 stars to look for possible evidence for accretion of planetary material.

\section{Observations}
\label{sec:observations}

The observations were carried out with the UVES spectrograph at the
VLT-UT2 Kueyen telescope. For TrES-1, OGLE-TR-10, 56, 111, and 113, the 
data were obtained between April and May 2005 in service mode 
(program ID\,075.C-0185). More details are presented in \citet[][]{Santos-2006a}. 

\begin{table*}
\caption{Stellar parameters and metallicities for the 6 stars studied in the current paper.}
\label{table:parameters}
\begin{tabular}{lcccrccl}
\hline\hline
Star & T$\mathrm{eff}$ [K] & $\log{g}$ (c.g.s.) & $\xi_{\mathrm{t}}$ [km\,s$^{-1}$] & \multicolumn{1}{c}{[Fe/H]} & N(\ion{Fe}{i}, \ion{Fe}{ii}) & $\sigma$(\ion{Fe}{i},\ion{Fe}{ii}) & Source\\
\hline
\object{OGLE-TR-10}  & 6075$\pm$86  & 4.54$\pm$0.15 & 1.45$\pm$0.14 & 0.28$\pm$0.10 & 33,11 & 0.08,0.06 & \citet[][]{Santos-2006a}\\
\object{OGLE-TR-56}  & 6119$\pm$62  & 4.21$\pm$0.19 & 1.48$\pm$0.11 & 0.25$\pm$0.08 & 31,9  & 0.06,0.08 & \citet[][]{Santos-2006a}\\
\object{OGLE-TR-111} & 5044$\pm$83  & 4.51$\pm$0.36 & 1.14$\pm$0.10 & 0.19$\pm$0.07 & 31,7  & 0.07,0.18 & \citet[][]{Santos-2006a}\\
\object{OGLE-TR-113} & 4804$\pm$106 & 4.52$\pm$0.26 & 0.90$\pm$0.18 & 0.15$\pm$0.10 & 30,5  & 0.10,0.09 & \citet[][]{Santos-2006a}\\
\object{OGLE-TR-132} & 6210$\pm$59  & 4.51$\pm$0.27 & 1.23$\pm$0.09 & 0.37$\pm$0.07 & 30,8  & 0.05,0.10 & Pont et al. (2006, in prep.)\\
\object{TrES-1}      & 5226$\pm$38  & 4.40$\pm$0.10 & 0.90$\pm$0.05 & 0.06$\pm$0.05 & 36,7  & 0.04,0.05 & \citet[][]{Santos-2006a}\\
\hline
\hline
\end{tabular}
\end{table*}

\begin{table}[] \centering
\caption{Line list used in the current paper for Na, Mg, Al, Si, Ca, Sc, Ti, V, 
Cr, Mn, Co, and Ni. The full table is available in electronic form at CDS.}
\begin{tabular}{ccc}	
\hline
\hline
$\lambda$ & $\chi_{l}$ & $\log{gf}$ \\
\hline
\ion{Si}{i}&&		      \\
5665.56 & 4.92 & -2.0000      \\
5690.43 & 4.93 & -1.8239      \\
5701.10 & 4.93 & -2.0372      \\
...     & ...  & ... \\
\hline
\hline
\end{tabular}
\label{table:lines}
\end{table}

For OGLE-TR-132 we have obtained 8 exposures of 3000 seconds 
each (program ID\,076.C-0131). The observations were done in December 2005 
and January 2006. Each individual spectrum
was then combined using the IRAF\footnote{IRAF is distributed by National 
Optical Astronomy Observatories, operated by the Association of Universities for
Research in Astronomy, Inc.,under contract with the National Science
Foundation, U.S.A.} {\tt scombine} routine. 
The total S/N obtained is close to 100, as measured directly from small
spectral windows with no clear spectral lines in 
the region near 6500\AA.
As for the other faint OGLE stars, for each 
exposure on OGLE-TR-132 the CCD was read in 2x2 bins to reduce the readout noise 
and increase the number of counts in each bin. This procedure 
does not compromise the resolving power, since the sampling of the CCD is still
higher (by a factor of 2) than the instrumental PSF.
Similarly we opted for using a slit
width of 0.9 arcsec, which provides a spectral resolution 
R=$\lambda$/$\Delta\lambda$$\sim$50\,000. The observations
were made using the Dichroic 390+580 mode. The red portion of the
spectra (used in the current paper) cover the wavelength domain 
between 4780 and 6805\AA, with a gap between 5730 and 5835\AA.

As before, particular attention was paid to the orientation of 
the slit due to the relative crowdedness of the field. The angle was chosen
using the images available at the OGLE 
website\footnote{http://www.astrouw.edu.pl/$\sim$ftp/ogle/index.html}, 
so that no other star was present in the UVES slit during the observation.

\section{Stellar parameters and iron abundances}
\label{sec:parameters}

Stellar parameters and iron abundances for our targets were taken from
\citet[][]{Santos-2006a} for TrES-1, OGLE-TR-10, 56, 111 
and 113, and from Pont et al. (2006, in prep.) for OGLE-TR-132.
These were derived in LTE using the 2002 version of the code MOOG 
\citep[][]{Sneden-1973}\footnote{http://verdi.as.utexas.edu/moog.html} 
and a grid of Kurucz Atlas plane-parallel model atmospheres \citep[][]{Kurucz-1993}. 
The whole procedure is described in \citet[][and references therein]{Santos-2004b},
and is based on the analysis of 39 \ion{Fe}{i} and 12 \ion{Fe}{ii} weak lines, and
imposing excitation and ionization equilibrium.

Details about the analysis of the 6 stars in our sample are presented in \citet[][]{Santos-2006a} and 
Pont et al. (2006, in prep.). The stellar parameters used in the current
paper are summarized in Table\,\ref{table:parameters}.

It is useful to mention that the stellar parameters used for the OGLE stars 
and TrES-1 were derived using the same methodology as
the ones used by \citet[][]{Bodaghee-2003}, \citet[][]{Gilli-2006}, and \citet[][]{Beirao-2005}
in their analysis of the same elements in the field star samples mentioned below. 
This gives us a guarantee of uniformity in the comparison presented in Sect.\,\ref{sec:galactic}.

\section{Abundances for other elements}
\label{sec:abundances}

The analysis was done in LTE 
using a grid of \citet[][]{Kurucz-1993} ATLAS-9
model atmospheres, and the 2002 version of the code MOOG \citep[][]{Sneden-1973}.

The abundances for Na, Mg, Al, Si, Ca, Sc, Ti, V, Cr, Mn, Co, and Ni were derived from
the analysis of line-equivalent widths (EW), measured by Gaussian fitting using
the {\tt splot} routine in IRAF. To derive the abundances we have followed the procedure
described in \citet[][]{Beirao-2005} for the first three elements
of this series, and in \citet[][]{Bodaghee-2003} and \citet[][]{Gilli-2006}
for the remaining cases.

\begin{table*}
\caption{Derived abundances for the alpha elements in the 6 stars studies in this paper. The errors denote the rms around 
the average abundance given by each of the n(X) lines used for the element X.}
\label{table:alpha}
\begin{tabular}{lcccccccccc}
\hline\hline
Star        &  [Si/H]         &n(Si)&  [Ca/H]          &n(Ca) &  [TiI/H]         & n(TiI)&  [TiII/H]      &n(TiII)& [Sc/H]       & n(Sc) \\
\hline
OGLE-TR-10  &  0.32$\pm$0.04  &  8  &  0.30$\pm$0.04   &  10  &  0.31$\pm$0.05   &  9	 &  0.31$\pm$0.09 &  4  &  0.33$\pm$0.10 &  4 \\
OGLE-TR-56  &  0.28$\pm$0.05  &  9  &  0.26$\pm$0.08   &  11  &  0.22$\pm$0.04   &  6	 &  0.29$\pm$0.05 &  4  &  0.35$\pm$0.05 &  6 \\
OGLE-TR-111 &  0.10$\pm$0.04  &  5  &  0.21$\pm$0.08   &  9   &  0.35$\pm$0.07   &  10   &  0.15$\pm$0.10 &  4  &  0.11$\pm$0.13 &  2 \\
OGLE-TR-113 &  0.16$\pm$0.12  &  7  &  0.00$\pm$0.07   &  8   &  0.29$\pm$0.09   &  13   &  0.20$\pm$0.17 &  4  &  0.14$\pm$0.12 &  4 \\
OGLE-TR-132 &  0.36$\pm$0.03  &  8  &  0.31$\pm$0.04   &  12  &  0.35$\pm$0.07   &  6	 &  0.36$\pm$0.10 &  4  &  0.49$\pm$0.03 &  7 \\
TrES-1      &  0.06$\pm$0.04  &  9  &  0.05$\pm$0.07   &  12  &  0.11$\pm$0.06   &  13   &  0.00$\pm$0.01 &  4  &  0.02$\pm$0.06 &  5 \\
\hline
\hline
\end{tabular}
\end{table*}

\begin{table*}
\caption{Same as Table\,\ref{table:alpha} for the iron-peek elements Mn, V, Cr and Co.}
\label{table:iron}
\begin{tabular}{lcccccccccc}
\hline\hline
Star        &  [Mn/H] &n(Mn)&  [V/H] &n(V) &  [CrI/H] & n(CrI) &  [CrII/H] &n(CrII)&  [Co/H] &n(Co) \\
\hline
OGLE-TR-10 &0.31$\pm$0.08 & 3 & 0.31$\pm$0.07 & 3 & 0.38$\pm$0.25 & 2 &   0.32$\pm$0.01 & 2 & 0.39$\pm$0.00 & 1   \\
OGLE-TR-56 &0.33$\pm$0.16 & 3 & 0.23$\pm$0.07 & 4 & 0.32$\pm$0.00 & 1 &   0.25$\pm$0.18 & 2 & 0.25$\pm$0.11 & 4   \\
OGLE-TR-111&0.37$\pm$0.13 & 3 & 0.69$\pm$0.14 & 8 & 0.21$\pm$0.09 & 4 &$-$0.02$\pm$0.00 & 1 & 0.30$\pm$0.10 & 8   \\
OGLE-TR-113&0.29$\pm$0.08 & 2 & 0.56$\pm$0.16 & 9 & 0.10$\pm$0.08 & 4 &   0.19$\pm$0.00 & 1 & 0.30$\pm$0.15 & 7   \\
OGLE-TR-132&0.37$\pm$0.12 & 3 & 0.32$\pm$0.18 & 4 & 0.25$\pm$0.03 & 2 &   0.30$\pm$0.10 & 3 & 0.34$\pm$0.09 & 6   \\
TrES-1     &0.14$\pm$0.01 & 3 & 0.21$\pm$0.07 & 8 & 0.04$\pm$0.02 & 4 &   0.00$\pm$0.00 & 1 & 0.06$\pm$0.09 & 7   \\
\hline
\hline
\end{tabular}
\end{table*}

\begin{table*}
\caption{Same as Table\,\ref{table:alpha} for the iron-peek element Ni, and for Na, Mg and Al.}
\label{table:namgal}
\begin{tabular}{lcccccccc}
\hline\hline
Star        &  [Ni/H] & n(Ni) &  [Na/H] &n(Na)&  [Mg/H] &n(Mg) &  [Al/H] & n(Al) \\
\hline
OGLE-TR-10  & 0.22$\pm$0.08 & 21 &   0.24$\pm$0.11 & 3 & 0.23$\pm$0.14 & 3 & 0.29$\pm$0.00 & 1 \\
OGLE-TR-56  & 0.27$\pm$0.06 & 23 &   0.38$\pm$0.09 & 3 & 0.22$\pm$0.17 & 3 & 0.30$\pm$0.01 & 2 \\
OGLE-TR-111 & 0.22$\pm$0.10 & 22 &   0.21$\pm$0.17 & 3 & 0.23$\pm$0.12 & 3 & 0.28$\pm$0.15 & 2 \\
OGLE-TR-113 & 0.17$\pm$0.10 & 20 &   0.06$\pm$0.17 & 3 & 0.22$\pm$0.18 & 3 & 0.22$\pm$0.02 & 2 \\
OGLE-TR-132 & 0.36$\pm$0.10 & 22 &   0.31$\pm$0.02 & 3 & 0.26$\pm$0.10 & 2 & 0.40$\pm$0.03 & 2 \\
TrES-1      & 0.06$\pm$0.04 & 26 &$-$0.05$\pm$0.06 & 3 & 0.05$\pm$0.08 & 3 & 0.10$\pm$0.01 & 2 \\
\hline
\hline
\end{tabular}
\end{table*}


\begin{table*}
\caption{Same as Table\,\ref{table:alpha} for C, O, S, Zn and Cu. For Cu, in OGLE-TR-113 
the abundance was obtained only from the 5380\,\AA\ line; the uncertainty was derived based 
on an estimate of the measured EW. For TrES-1 only the 6757\,\AA\ line of sulphur was measured.}
\label{table:ecuvillon}
\begin{tabular}{lccccc}
\hline\hline
Star        &  [C/H]         & [O/H]     &  [S/H]  & [Zn/H]       & [Cu/H] \\
\hline
OGLE-TR-10  & 0.29$\pm$0.09   &     0.31$\pm$0.20&	   0.13$\pm$0.12&	0.47$\pm$0.11 &     0.27$\pm$0.10  \\
OGLE-TR-56  & 0.24$\pm$0.14   &     0.35$\pm$0.26&	   0.20$\pm$0.09&	0.09$\pm$0.09 &     0.23$\pm$0.08  \\
OGLE-TR-111 & 0.43$\pm$0.21   &     0.21$\pm$0.38&	$<$0.64 	&	0.13$\pm$0.11 &     0.47$\pm$0.10  \\
OGLE-TR-113 & 0.81$\pm$0.21   &     -		 &	$<$0.70 	&	0.34$\pm$0.11 &     0.53$\pm$0.11  \\
OGLE-TR-132 & 0.27$\pm$0.10   &     -		 &	   0.42$\pm$0.11&	0.21$\pm$0.09 &     0.30$\pm$0.09  \\
TrES-1      & 0.06$\pm$0.07   &     0.11$\pm$0.17&	$<$0.26 	&	0.05$\pm$0.06 &     0.04$\pm$0.08  \\
\hline
\hline
\end{tabular}
\end{table*}

For the 12 elements listed above, 
we used the {\tt abfind} driver in MOOG to find the abundances for each 
measured line giving as input the model atmosphere interpolated to
the correct stellar parameters listed in Table\,\ref{table:parameters},
the measured EW, and the atomic parameters provided in Table\,\ref{table:lines}.
The line-list used is an upgraded version of the one used in
\citet[][]{Gilli-2006}, where we added a few 
more lines (except for Na, Mg and Al). Semi-empirical $\log{gf}$ values were derived, as before,
using equivalent widths obtained in the High Resolution Solar Atlas
\citep[][]{Kurucz-1984} and a solar model atmosphere with 
T$_{\mathrm{eff}}$$=$5777\,K, $\log{g}$$=$4.44\,dex, and $\xi_t$$=$1.00\,km\,s$^{-1}$.
The solar abundances were taken from \citet{Anders-1989}, except for iron,
where the value of $\log{\epsilon}\mathrm{(Fe)}$ $=$7.47 was considered,
as in all our previous studies. In our analysis of Na, Mg, Al, Si, Ca, Sc, 
Ti, V, Cr, Mn, Co, and Ni we used the damping option ``2'' in MOOG.

For Cr and Ti we used both neutral (\ion{Cr}{i} and \ion{Ti}{i}) and 
ionized (\ion{Cr}{ii} and \ion{Ti}{ii}) lines. For the rest of the
comparison we will use the abundances derived using the neutral lines,
since the studies about the field stars were based only on the neutral species. 
A look at Tables\,\ref{table:alpha} and \ref{table:iron} shows that
the two sets of lines yield generally similar results, within the errors.

Oxygen (O) abundances were derived from the forbiden [OI] line near 6300.3\,AA.
We followed the analysis procedure described in detail in \citet[][]{Ecuvillon-2006}.
Given that in metal-rich stars the [OI] line is blended with a small nickel 
line \citep[e.g.][]{Allende-2001}, in our analysis we have computed the expected EW for 
this latter line, and subtracted if from the whole EW of the feature at 6300.3\,\AA.
For OGLE-TR-113 and OGLE-TR-132, the quality of the spectra in the [OI] line
region did not permit us to derive a quality value for the oxygen abundance.
As in \citet[][]{Ecuvillon-2006}, we adopted a solar abundance 
of $\log{\epsilon}\mathrm{(O)}$$=$8.56. As for C, S, Zn and Cu (below),
in this case we adopted the damping option ``0'' in MOOG (Unsold approximation). 

\begin{figure}[t!]
\resizebox{\hsize}{!}{\includegraphics{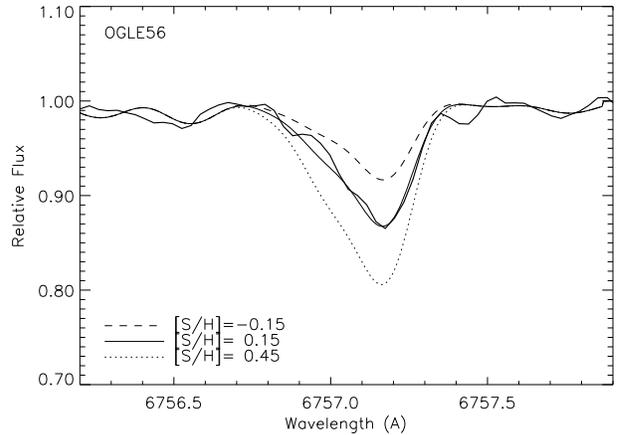}}
\caption{Spectral synthesis of the S line at 6757.1\,\AA\ for OGLE-TR-56. The solid
line represents the observed spectrum, while three fits are denoted by the thinner lines.}
\label{fig:synthS}
\end{figure}

\begin{figure}[t!]
\resizebox{\hsize}{!}{\includegraphics{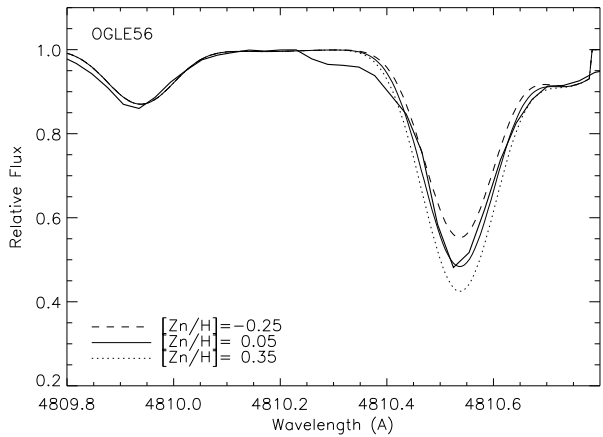}}
\caption{Spectral synthesis of the Zn line at 4810.5\,\AA\ for OGLE-TR-56.}
\label{fig:synthZn}
\end{figure}

\begin{figure}[t!]
\resizebox{\hsize}{!}{\includegraphics{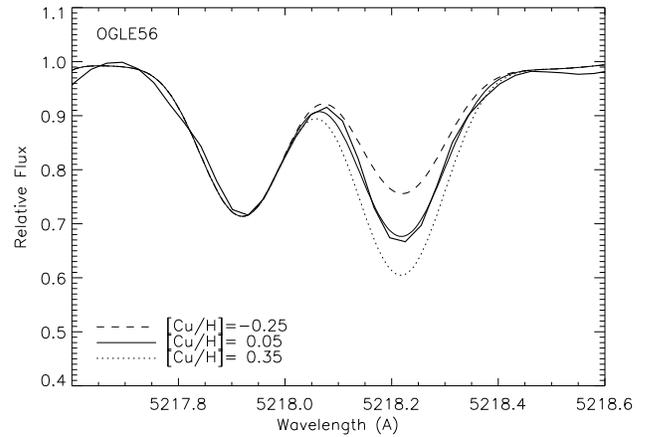}}
\caption{Spectral synthesis of the Cu line at 5218.2\,\AA\ for OGLE-TR-56.}
\label{fig:synthCu}
\end{figure}

For the abundances of carbon (C), sulphur (S), zinc (Zn) and copper (Cu) we 
followed the prescription of \citet[][]{Ecuvillon-2004b}. Carbon abundances 
were derived using an EW analysis of two \ion{C}{i} lines at 5380.3 
and 5052.2\,\AA. Sulphur, zinc and copper abundances were all obtained by 
spectral fitting to the data, using the {\tt synth} driver in MOOG. For S 
we have used two \ion{S}{i} features near 6743.5 and 6757.1\,\AA. Zinc 
abundances were obtained from the \ion{Zn}{i} line at 4810.5\,\AA, since the shorter
wavelenth line at 4722.2\,\AA\ used by \citet[][]{Ecuvillon-2004b} is
not present in our spectra. Finally, Cu abundances were derived from the
analysis of the \ion{Cu}{i} 5218.2\,\AA\ line, since the feature at 5782.1\,\AA\
falls in the spectral gap of the UVES spectra. 
In the spectral synthesis, we considered a gaussian broadening to take into account 
the instrumental profile. When available from the literature 
\citep[][]{Alonso-2004,Pont-2004,Bouchy-2004,Bouchy-2005a}, the projected rotational velocity $v\,\sin{i}$ was
also considered. When only an upper limit for the $v\,\sin{i}$ was available, we made use of other lines in 
the same spectral region to optimize the broadening function, following the same procedure used in
\citet[][]{Ecuvillon-2004b}.

We refer to \citet[][]{Ecuvillon-2004b} and \citet[][]{Ecuvillon-2006} for
the line-lists used and for more detail on the the derivation of the abundances 
of O, C, S, Zn and Cu. An example of the spectral synthesis for the derivation
of S, Zn and Cu abundances in presented in Figs.\,\ref{fig:synthS}, \ref{fig:synthZn} 
and \ref{fig:synthCu}, respectively.

\begin{table*}[t!]
\caption{Stellar I magnitudes, stellar radii, galactic coordinates,
derived distances, X, Y, Z, R$_g$ galactic coordinates, adopted interstellar extinction, 
and systemic radial-velocities for the 6 stars studied in this paper. The distance derived considering null
extition, d$_0$, is also presented. For TrES-1, the I magnitude was derived from
the V magnitude (V=11.79) and the known effective temperature. See text for more details.
}
\label{table:galactic}
\begin{tabular}{lcccccccccccc}
\hline\hline
Star &  I       & R$_{\mathrm{star}}$ & $l$    & $b$    & d$_0$ & d   &  X   & Y    & Z    & R$_g$ & Av     & RV \\
     &  [mag.]  & [R$_{\odot}$]       & [$^o$] & [$^o$] & [pc]  &[pc] & [pc] & [pc] & [pc] & [pc]  & [mag.] & [km\,s$^{-1}$]\\
\hline
OGLE-TR-10  &  14.9  &1.14  & 359.8516  &-1.5767 &  1830  &  1326  &	-3  &  7174  &   -36  &  7175  &  1.16 & $-$6.2$^a$\\
OGLE-TR-56  &  15.3  &1.15  &	0.7053  &-2.3655 &  2246  &  1591  &	20  &  6911  &   -66  &  6911  &  1.25 & $-$48.3$^a$\\
OGLE-TR-111 &  15.5  &0.83  & 289.2794  &-1.7056 &  1236  &  1011  &  -954  &  8166  &   -30  &  8222  &  0.72 & 25.1$^b$\\
OGLE-TR-113 &  14.4  &0.765 & 289.2017  &-1.7895 &   614  &   553  &  -522  &  8318  &   -17  &  8335  &  0.42 & $-$7.9$^c$\\
OGLE-TR-132 &  15.7  &1.28  & 289.2349  &-2.3430 &  3081  &  2180  & -2057  &  7782  &   -89  &  8050  &  1.25 & 39.7$^c$\\
TrES-1      &  10.73 &0.83  &  67.4649  &13.4403 &   148  &   143  &   129  &  8447  &    33  &  8448  &  0.11 & $-$20.7$^d$\\
\hline
\hline
\end{tabular}
\newline
$^a$ \citet[][]{Bouchy-2005a}; $^b$ \citet[][]{Pont-2004}; $^c$ \citet[][]{Bouchy-2004}; $^d$ This paper
\end{table*}

The final derived abundances are listed in Tables\,\ref{table:alpha}, 
\ref{table:iron}, \ref{table:namgal}, and \ref{table:ecuvillon}. 
In the first three, the uncertainties represent the rms around the average abundance for 
the cases where more than one line of the same element could be measured. 
For C, O, S, Zn, and Cu the final errors were computed adding in quadrature the 
errors due to the uncertainties in atmospheric parameters following the sensitivities listed in
\citet[][]{Ecuvillon-2004b} and \citet[][]{Ecuvillon-2006}, together with the uncertainty in the abundance
synthesis or analysis. For this latter, the rms was used when more than one line was measured. Else, 
the uncertainty in the measured EW was considered. For S, Zn, and Cu the uncertainty in the synthesis procedure 
was taken into account.

\subsection{Comparison with the literature}
\label{sec:comparison}

The chemical abundances for the faint OGLE stars have not been studied
elsewhere in the literature. However, for the brighter TrES-1, \citet[][]{Sozzetti-2006}
have derived chemical abundances for a series of elements, most of which
are also studied in the current paper.



In general, the abundances derived by \citet[][]{Sozzetti-2006} for TrES-1 are in agreement with the ones
presented here. However, we do see a systematic difference between the two
studies. On average, our values are slighly above (by 0.08\,dex, with a dispersion
around the average value of 0.08) the ones derived
by the study of Sozzetti et al. This small systematic difference can be due to the use 
of slighly different atmospheric parameters and line-lists. 

\section{Distances and galactic positions}
\label{sec:distances}

We used the available information regarding the 6 stars in
our sample to derive their distances and galactic positions. 

Since most of the targets (all the OGLE stars) have only I magnitudes available, we decided
to work in this band, and derived the I magnitude of TrES-1 
by inverting the T$_{\mathrm{eff}}$:(V-I) calibration of \citet[][]{Alonso-1996}. For this we
considered V=11.79 and made use of the effective temperature listed in Table\,\ref{table:parameters}.

Using the effective temperatures listed in the Table and the best available stellar radii estimates 
\citep[Pont et al. 2006, in prep.;][]{Bouchy-2004,Laughlin-2005,Santos-2006a}
we derived the stellar luminosities. We then used the M$_{\mathrm{bol}}$:L relation presented 
in \citet[][]{Lang-1999} to obtain the bolometric magnitude of each star.
From M$_{\mathrm{bol}}$, the absolute magnitudes (M$_{\mathrm{I}}$) were derived
using the Bolometric Correction for the I band (BC$_{\mathrm{I}}$) taken from \citet[][]{Bessel-1998}.
Finally, with M$_{\mathrm{I}}$ and the apparent I magnitude, and neglecting the
interstellar absorption, we could derive a first guess for the stellar distances 
(d$_\mathrm{0}$).

\begin{figure}[b!]
\resizebox{\hsize}{!}{\includegraphics[bb=120 160 525 685,clip]{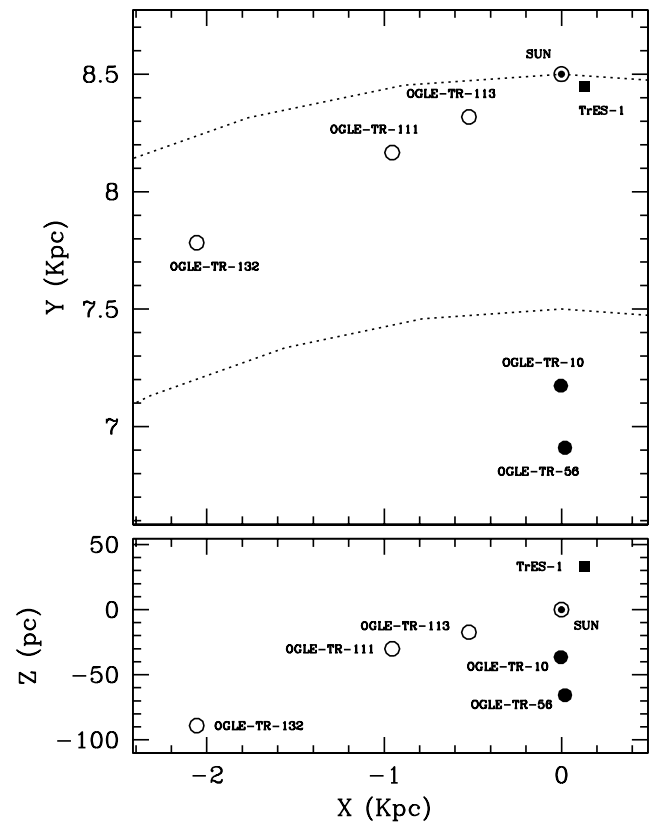}}
\caption{Galactic cartesian X, Y and Z coordinates of our stars and the Sun. The 
coordinate system is right-handed. Here we 
considered that the Sun is at a galactocentric radius of 8.5\,Kpc, in the Y direction. 
Different symbols are considered for stars in the different fields: OGLE-TR-10 and 56 
(filled circles), OGLE-TR-111, 113 and 132 (open circles), and TrES-1 
(filled square). The dotted curves represent the location of the
8.5 and 7.5\,Kpc galactocentric regions. In this figure we 
are basically representing the 4th galactic quadrant.}
\label{fig:distances}
\end{figure}

Once a first value was available, we used a fortran version of the IDL code {\tt EXTIN.PRO} 
\citep[][]{Amores-2005}\footnote{The code was kindly provided by the authors. The IDL version is available at http://www.astro.iag.usp.br/$\sim$jacques/programs.html} and the 
relation A$_{\mathrm{I}}$/A$_{\mathrm{V}}$$\sim$0.601 \citep[][]{Schlegel-1998}\footnote{Similar results are derived
from the tables of \citet[][]{Grebel-1995}.}
to obtain a value for A$_{\mathrm{V}}$ and A$_{\mathrm{I}}$. This latter was then used to obtain
a new value for the stellar distance. All this procedure was done iteractively, until the distances remained
unchanged. The final values, as well as the first guess (considering null extinction), 
are presented in Table\,\ref{table:galactic}. The null-extinction value for TrES-1 
is very close to the one derived in previous works \citep[][]{Alonso-2004,Laughlin-2005}.
For this star, using the observed V magnitude, the bolometric correction from \citet[][]{Flower-1996},
and considering zero reddening, a distance of 158\,pc is derived.




Having the distance to the stars, and the {\it l, b} galactic coordinates\footnote{Derived using the
epoch 2000.0 right-ascention and declinations, and making use of the the {\tt galactic} routine within the IRAF ``astutils'' package.}
we obtained the X, Y and Z galactic positions for the 6 stars (Fig.\,\ref{fig:distances}, 
Table\,\ref{table:galactic}). As we can see from the figure, the OGLE stars in
the different fields (Carina at $l\sim289$ degrees and Bulge at $l\sim0$ degrees)
are all at very different distances from the Sun. Althought the Carina-field 
stars (OGLE-TR-111, 113 and 132) have all a similar galactocentric radius 
as the Sun, the same is not true for the Bulge-field stars (OGLE-TR-10 and 56).
Here we consider that the Sun is at R$_g$=8.5\,Kpc.

\begin{figure*}[t!]
\resizebox{\hsize}{!}{\includegraphics{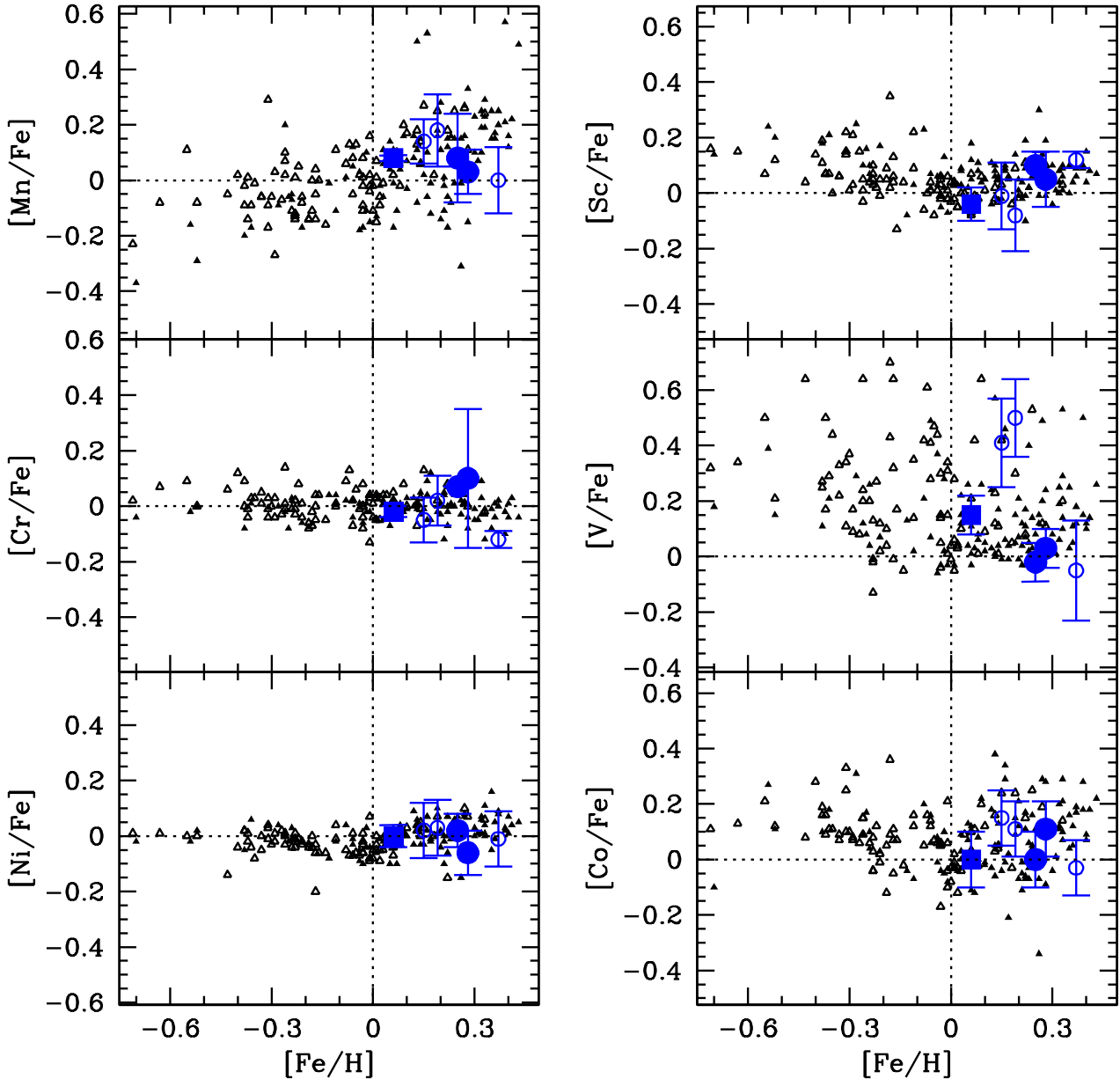}}
\caption{[X/Fe] vs. [Fe/H] plots for X = Ni, Cr, Mn, Co, V (Fe-peek elements) 
and Sc (alpha-element). Filled triangles represent stars with giant planets, and open triangles 
denote single field stars. The other symbols represent OGLE-TR-10 and 56 (filled circles),
OGLE-TR-111, 113 and 132 (open circles), and TrES-1 (filled square). The error bars represent 
the rms around the average abundance when more than one line was used to derive the chemical 
abundances. When one single line was used, we adopted an error of 0.10\,dex in these plots.}
\label{fig:abundances1}
\end{figure*}

\begin{figure*}[t!]
\resizebox{\hsize}{!}{\includegraphics{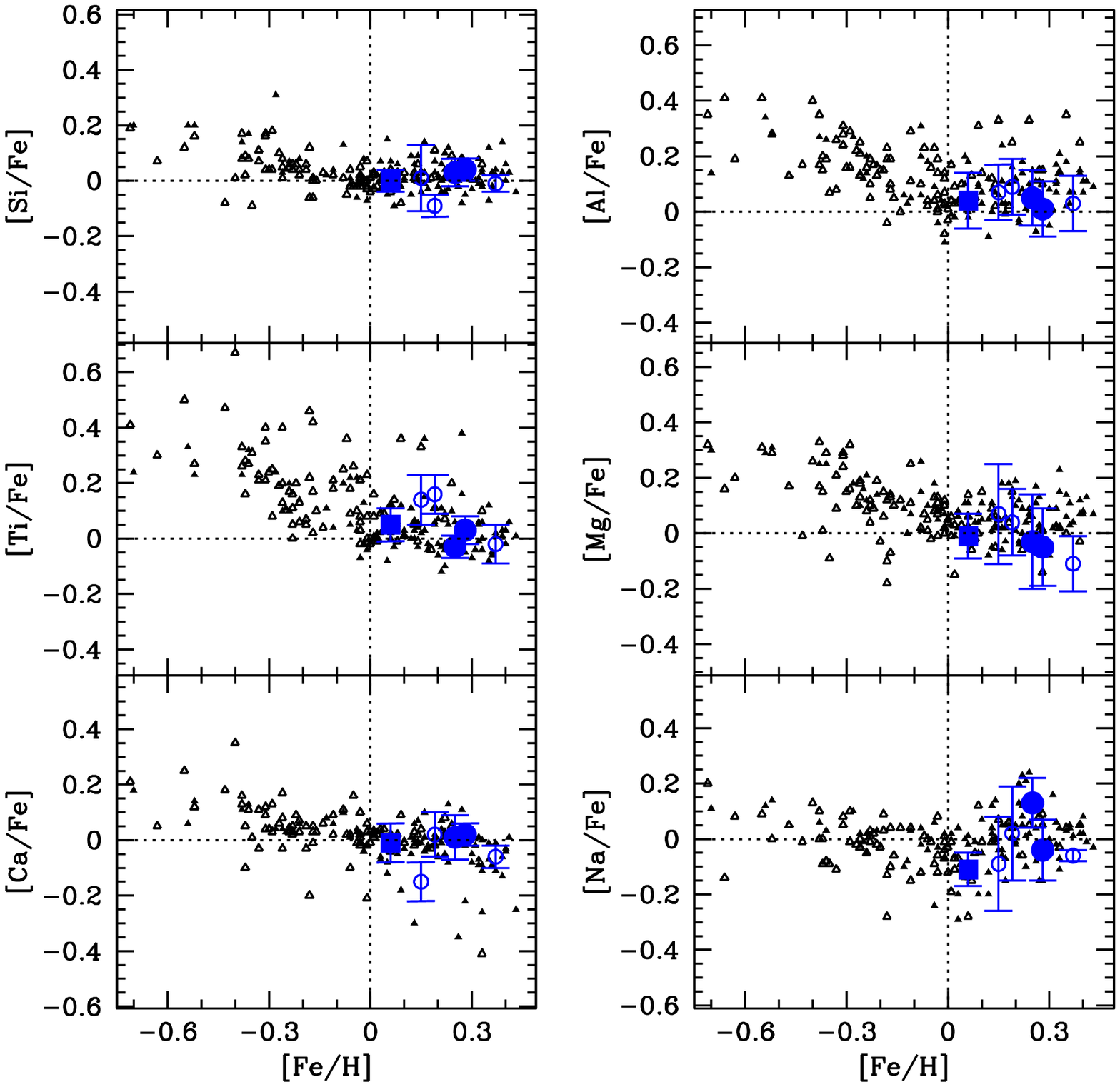}}
\caption{Same as Fig.\,\ref{fig:abundances1} for X = Ca, Ti, Si (alpha-elements), Na, Mg and Al.}
\label{fig:abundances2}
\end{figure*}

We should add that errors in the stellar radii, magnitudes, and temperatures, together
with possible large uncertainties in the determination of the interstellar
reddening, imply that the values listed in Table\,\ref{table:galactic} must be seen as
indicative.
	

\section{Galactic chemical trends}
\label{sec:galactic}

{ The use of chemical abundances to distinguish between different populations 
in the Galaxy (in particular the thin disk, thick disk, and galactic bulge) has been thoroughly discussed 
in the literature \citep[e.g.][]{Bensby-2003,Nissen-2004,Fuhrmann-2004,Brewer-2006,Fulbright-2005}. The 
observed differences are though to reflect different star formation histories in the different
galactic systems.} It is thus interesting to check if
the abundances of the elements studied in the current paper for TrES-1 
and the 5 OGLE stars differ or not from the ones measured in field disk 
stars.

\begin{figure*}[t!]
\resizebox{\hsize}{!}{\includegraphics[bb=10 230 550 660,clip]{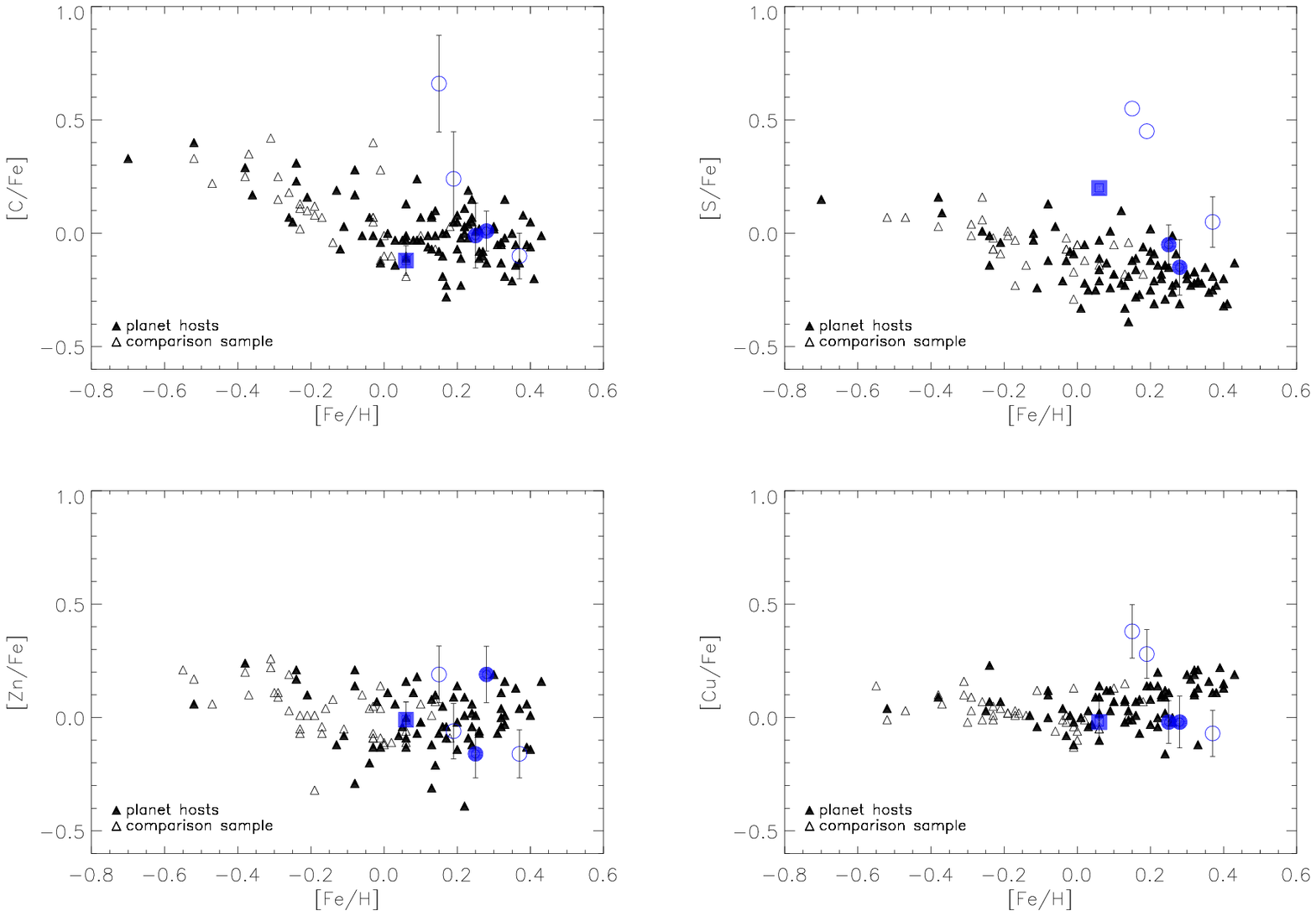}}
\caption{Same as Fig.\,\ref{fig:abundances1} for X=C, S, Zn, and Cu. For S, the points of
TrES-1, OGLE-TR-111 and 113 (without error bars) represent upper limits in the abundances.}
\label{fig:CSZnCu}
\end{figure*}

\begin{figure}[t!]
\resizebox{\hsize}{!}{\includegraphics{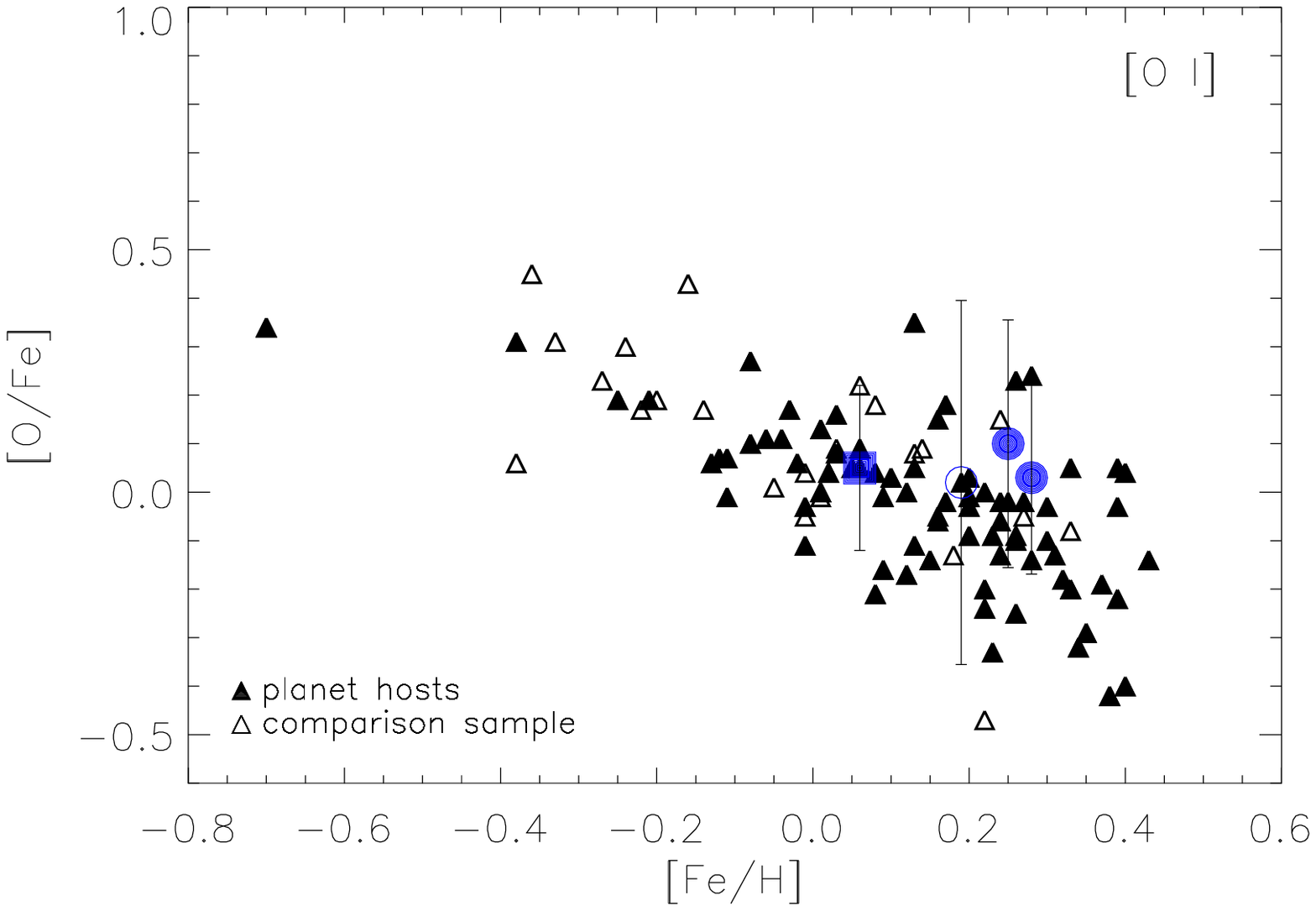}}
\caption{Same as Fig.\,\ref{fig:abundances1} for Oxygen.}
\label{fig:oxygen}
\end{figure}

In Figs.\,\ref{fig:abundances1}, \ref{fig:abundances2}, \ref{fig:CSZnCu} and \ref{fig:oxygen}
we present the [X/Fe] vs. [Fe/H] plots for all the elements studied in the current paper.
In the figures, the small triangles denote the solar-neighborhood field stars
studied in \citet[][]{Bodaghee-2003}, \citet[][]{Beirao-2005}, and \citet[][]{Gilli-2006} 
(cases of Na, Mg, Al, Si, Ca, Sc, Ti, V, Cr, Mn, Co, Ni), and in \citet[][]{Ecuvillon-2004b,Ecuvillon-2006} 
(for O, C, S, Cu and Zn). Filled triangles correspond to stars
with giant planets, and open triangles denote ``single'' (without known planets) field stars. The choice of
these samples is based on the fact that these were studied using the same line-lists, analysis 
methods, and model atmospheres used in the current study, and thus give us confidence that
a relative comparison will be unbiased.
The 6 stars studied in this paper are denoted by the larger symbols. 
For the 3 Carina field stars (OGLE-TR-111, 113 and 132) we used an open circle, for the 2 bulge 
targets (OGLE-TR-10 and 56) we used a filled circle, while TrES-1 is denoted by a filled square. 
The choice of different symbols is important since these stars belong to different populations 
in the Galaxy (see Sect.\,\ref{sec:distances}).

{ As we can see from the plots, the 5 OGLE stars and TrES-1 are in a first analysis 
chemically indistinguishable from the field stars used as reference (mostly thin disk stars -- 
Ecuvillon et al., in preparation), 
except for the fact that they represent a particularly metal-rich ``population''. No major differences 
are found for any of the studied elements. In particular, no important alpha-element 
enhancement is observed. Such a trend could be typical of galactic bulge stars \citep[e.g.][]{Fulbright-2005} or 
of thick disk objects \citep[e.g.][]{Fuhrmann-2004}.
The radial-velocities presented in Table\,\ref{table:galactic} also do not show any particular anomaly.}
These same conclusions were also taken by \citet[][]{Sozzetti-2006} concerning TrES-1. 

For the closest stars in the Carina field
(OGLE-TR-111, 113), all with galactocentric radii similar to
the Sun, this result is probably not unexpected. However, some differences
could be expected for OGLE-TR-10 and 56, located at $\sim$1.5 Kpc in the
direction of the galactic center, and for OGLE-TR-132, at more than 2 Kpc from us. 
The fact that the studied stars are not young \citep[][]{Melo-2006}, together with
the stellar galactic velocity dispertion \citep[][]{Nordstrom-2004}, may help to
explain this similarity.


A carefull look at the plots shows, however, that some small differences may exist.
For instance, OGLE-TR-132 seems to generally occupy a position in the lower envelope 
of the points in the field star distribution. Given that this systematic effect is observed for most
of the elements, this difference, although small, seems to be real. Interestingly, 
according to the calculations in Sect.\,\ref{sec:distances}, this star is the most
distant of all the targets studied here.

Important differences may also be found for OGLE-TR-111 and 113 regarding
V, C, Cu, and maybe Ti, where these two stars seem to be particularly overabundant.
However, we should mention that the observed differences for these elements
may be due to analysis effects. OGLE-TR-111 and 113 are the two coolest stars
in our sample. As shown by \citet[][]{Bodaghee-2003}, V and Ti are amongst the elements
presenting the strongest [X/Fe] vs. T$_{\mathrm{eff}}$ trends, possibly due to NLTE effects,
where the lowest temperature stars present higher abundances. 
The high abundances of C and Cu are less simple to explain, as no clear dependence
of the derived abundances of these two elements with effective temperature was found in the study
of \citet[][]{Ecuvillon-2004b}. The very high carbon abundance of OGLE-TR-113
is particularly intriguing, although the derived value is based only on
one of the two available carbon lines. 

The abundances of Cu for TrES-1, OGLE-TR-10 and 56 are
a bit below the field star sample, although in agreement within the errors.
A similar residual low value is found for Al and most clearly for Mg in these stars. 

Finally, the Ca abundance for OGLE-TR-113 seems to be particularly low, although the
other alpha elements do not present the same tendency.

\section{Condensation temperature}
\label{sec:tcond}

The accretion of planetary material has been proposed by several authors
to explain the high metal content observed in stars orbited by giant planets 
\citep[e.g.][]{Gonzalez-1998,Murray-2002}. This effect could be particularly 
important concerning the stars with very short period planets, since planetary
migration could increase the quantity of planetesimals falling into the
stellar surface \citep[][]{Murray-1998}.

Abundance traces of pollution events in the 
atmospheres of stars with giant planets have indeed been
reported in the literature in some particular cases \citep[][]{Israelian-2001,Laws-2001}. 
Although several observational results have made this hypothesis 
unlikely \citep[e.g.][]{Pinsonneault-2001,Santos-2003,Fischer-2005}, even if they did not 
completely discard it \citep[e.g.][]{Vauclair-2004}, it is interesting to explore this
possibility in our case, since all the 6 stars in our sample are both metal-rich and
orbited by short period giant planets.

If a solar-type stars engulfs a significant amount of planetary like material,
we can expect to observe that some elements have been enriched more
than others. Such an effect is expected since lower condensation temperature elements will probably 
evaporate before falling into the star. This idea may be valid, however, only if the infall
of material is slow enough, or if the infalling bodies are not too big, so that the low 
condensation temperature material is able to escape.
Having this possibility in mind, several authors
have studied the trends of element abundance as a function of the condensation temperature
of the elements in stars with giant planets \citep[][]{Smith-2001,Takeda-2001,Sadakane-2002,Ecuvillon-2006b}.

\begin{figure}
\resizebox{\hsize}{!}{\includegraphics{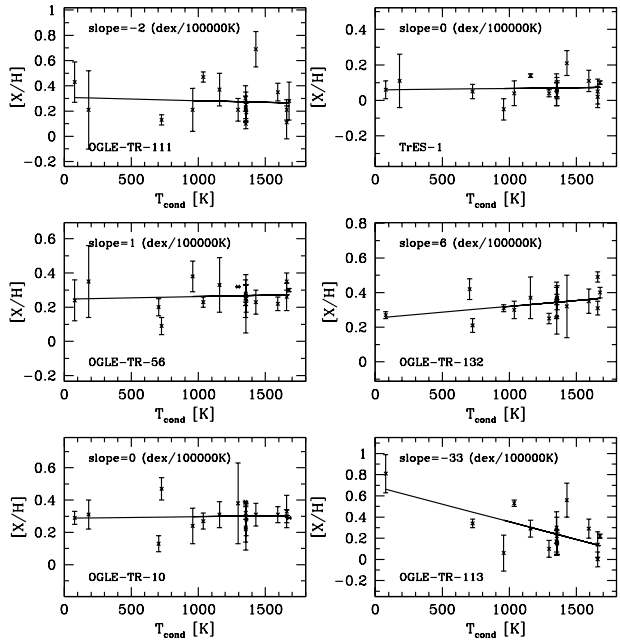}}
\caption{Element abundances as a function of the condensation temperature for the
6 stars studied in the current paper. The measured slopes are shown in each panel.}
\label{fig:tcond}
\end{figure}

Using the chemical abundances listed in Tables\,\ref{table:parameters} through \ref{table:ecuvillon} we
have derived the slopes of the relation between [X/H] and T$_{cond}$ for the stars in our sample (Fig.\,\ref{fig:tcond}).
We used the same methodology as in \citet[][]{Ecuvillon-2006b}, making use the condensation temperatures
for the elements listed in \citet[][]{Lodders-2003}. In our case, these go from 78\,K for carbon to
1677\,K for aluminum.

The derived T$_{cond}$-slopes are all between $-$2 (for OGLE-TR-111) and 6\,dex/100\,000\,K 
(for OGLE-TR-132), well within the values obtained for the other stars in 
the \citet[][]{Ecuvillon-2006b} sample. The only exception is OGLE-TR-113, 
whose T$_{cond}$-slope of $-$33\,dex/100\,000\,K
is due to the very high C abundance observed for this star. Excluding the carbon abundances for
this star, the slope obtained is only of $-$5\,dex/100\,000\,K. Unfortunately, for this star
the abundance of carbon is based on one single carbon line, and we cannot confirm
the observed C abundance. We note, however, that a
negative slope indicates a low refractory-to-volatile ratio, contrary to what would be expected
if pollution were responsible for any element abundance changes.

In all, we do not find any clear evidence for differential accretion in the 6 planet-host 
stars analyzed.

\section{Concluding remarks}

We have derived the abundances of C, O, Na, Mg, Al, Si, S, Ca, Sc, Ti, V, Cr, Mn, 
Co, Ni, Cu and Zn in the transiting planet-host stars OGLE-TR-10, 56, 111, 
113, 132 and TrES-1. Using the available planetary radii and effective temperatures,
as well as a model for the interstellar extition, we have also derived the
galactic positions of the 6 stars. The results show that OGLE-TR-10 and 56
are more than 1\,Kpc inside the solar radius, while the remaining stars
have similar galactocentric radius as the Sun, although at a
variety of distances. 

A comparison of the chemical abundances with the ones found for stars in the 
solar neighborhood shows that appart from the fact that they are particularly metal-rich,
no major differences are found regarding the relative element-over-iron abundances. 
Given their galactic positions, this result is probably not unexpected for OGLE-TR-111, 
113 and TrES-1, but may be particularly interesting for the cases 
of OGLE-TR-10, 56 and 132. 

Using the obtained chemical abundances and the condensation temperatures of
the elements, we have then explored the possibility that the 6 stars, all
orbited by short period giant planets, may have accreted planetary-like material.
The results show no clear evidence for such phenomena, giving stronger
support to the idea that the observed metal-content in the stars with giant
planets is not due to the accretion of metal-rich, hydrogen poor material.

\begin{acknowledgements}
   We would like to thank A. Moitinho, S. Sousa, J. Lepine and E. Amores 
   for the help in deriving the interstellar extinction values.
   Support from the Funda\c{c}\~ao para a Ci\^encia e a Tecnologia (Portugal)
   to N.C.S. in the form of a fellowship (reference SFRH/BPD/8116/2002)
   and a grant (reference POCI/CTE-AST/56453/2004) is gratefully
   acknowledged.
\end{acknowledgements}

\bibliographystyle{aa}
\bibliography{santos_bibliography}

\end{document}